\documentclass[aps,reprint,prd,showpacs,nofootinbib,twocolumn]{revtex4}

\usepackage{amsmath,amssymb}
\usepackage{graphicx,subfigure}
\usepackage{color,multirow}
\usepackage[colorlinks,linkcolor=magenta,anchorcolor=cyan,citecolor=blue,plainpages=false]{hyperref}

\hypersetup{colorlinks=true,
    breaklinks=true,
    pdfstartview=Fit,
    linkcolor=blue,
    citecolor=blue,
    urlcolor=blue}

\def\be{\begin{equation}}
    \def\ee{\end{equation}}
\def\ba{\begin{eqnarray}}
    \def\ea{\end{eqnarray}}

\begin{document}

\title{Dark energy in light of recent DESI BAO and Hubble tension}

\author{Hao Wang$^{1,2} $\footnote{\href{wanghao187@mails.ucas.ac.cn}{wanghao187@mails.ucas.ac.cn}}}
\author{Yun-Song Piao$^{1,2,3,4} $ \footnote{\href{yspiao@ucas.ac.cn}{yspiao@ucas.ac.cn}}}

    \affiliation{$^1$ School of Fundamental Physics and Mathematical
        Sciences, Hangzhou Institute for Advanced Study, UCAS, Hangzhou
        310024, China}

    \affiliation{$^2$ School of Physics Sciences, University of
        Chinese Academy of Sciences, Beijing 100049, China}

    \affiliation{$^3$ International Center for Theoretical Physics
        Asia-Pacific, Beijing/Hangzhou, China}

    \affiliation{$^4$ Institute of Theoretical Physics, Chinese
        Academy of Sciences, P.O. Box 2735, Beijing 100190, China}

    \begin{abstract}

Recently, Dark Energy Spectroscopic Instrument (DESI)
collaboration based on their first year data has reported a
$\gtrsim 3\sigma$ evidence for an evolving dark energy (DE)
against the cosmological constant (CC), so the standard
$\Lambda$CDM model. However, it is necessary to access the impact
of DESI data on the state equation $w_0$-$w_a$ of DE in the
Hubble-tension-free cosmologies, where $w_0$ and $w_a$ is the
parameters of state equation of DE. In this paper, using recent
DESI BAO measurements combined with Planck CMB and Pantheon Plus
dataset, we perform the Monte Carlo Markov Chain (MCMC) analysis
for the $w_0w_a$CDM model with possible pre-recombination
resolutions of the Hubble tension. It is found that though
$w_0>-1$ and $w_a<0$ are still preferred, the CC is also
$<2\sigma$ consistent, while the bestfit Hubble constant $H_0$ are
higher than those with pre-DESI BAO data but without the further
exacerbation of $S_8$ tension. According to our results, the
resolutions of Hubble tension are likely to suppress the
\textit{preference} of DESI for the evolving DE, thus the claim of
ruling out the CC needs to be more cautious regarding not only the
recent observational data but also the cosmological tensions.

    \end{abstract}

    \maketitle

\section{INTRODUCTION}

The standard $\Lambda$CDM model is thought to be the most
successful model to explain the majority of the cosmological
observations. Despite its huge success, currently it also suffers
from some serious issues, in particular the well-known Hubble
tension \cite{Verde:2019ivm,Riess:2019qba}, i.e. the SH0ES
collaboration using Cepheid-calibrated Type Ia supernovae measured
the Hubble constant $H_0\sim 73$km/s/Mpc \cite{Breuval:2024lsv},
while based on the $\Lambda$CDM model the Planck collaboration
inferred $H_0\sim 67$km/s/Mpc using the observations of cosmic
microwave background (CMB) \cite{Planck:2018vyg}.

The possibilities of solving the Hubble tension have been widely
studied, see
\cite{Knox:2019rjx,Perivolaropoulos:2021jda,DiValentino:2021izs,Vagnozzi:2023nrq}
for recent reviews. As one of the early (pre-recombination)
resolutions to the Hubble tension, early dark energy (EDE)
\cite{Karwal:2016vyq,Poulin:2018cxd,Smith:2019ihp} might be the
most competitive. In corresponding model, EDE field is not
negligible just before recombination, which suppressed the sound
horizon and consequently rised $H_0$. In particular an anti-de
Sitter phase around recombination can efficiently enhance the EDE
contribution (AdS-EDE) \cite{Ye:2020btb,Ye:2020oix}.
Though the rise of $H_0$ is usually accompanied with the
exacerbation of the $S_8$ tension (existed in $\Lambda$CDM)
\cite{Hill:2020osr,Ivanov:2020ril,DAmico:2020ods,Krishnan:2020obg,Nunes:2021ipq},
where $S_8=\sigma_8(\Omega_M/0.3)^{1/2}$ and $\sigma_8$ is the
amplitude of matter perturbations at $8h^{-1}$Mpc scale, it might
be alleviated by mechanisms independent of EDE
\cite{Poulin:2022sgp,Allali:2021azp,Ye:2021iwa,Alexander:2022own,FrancoAbellan:2021sxk,Clark:2021hlo,Simon:2022ftd,Reeves:2022aoi,Wang:2022bmk,Chacko:2016kgg,Buen-Abad:2022kgf,Teng:2021cvy,Dainotti:2021pqg,Dainotti:2022bzg,Montani:2024pou}.

In standard $\Lambda$CDM model, $\Lambda$ is a positive
cosmological constant (CC), which is also regarded as the dark
energy (DE), responsible for the accelerated expansion of our
current Universe. In the past decades, identifying the nature of
DE has been still an important challenge, see
e.g.\cite{Carroll:2000fy,Frieman:2008sn,Nojiri:2017ncd,Huterer:2017buf}
for reviews. However, recently using their first year data the
DESI collaboration \cite{DESI:2024lzq,DESI:2024mwx,DESI:2024uvr}
has reported a $\gtrsim 3\sigma$ evidence for an evolving DE
against the CC, so the standard $\Lambda$CDM model.
The DESI collaboration found, based on the $w_0w_a$CDM model where
the equation of state of DE satisfies
\cite{Chevallier:2000qy,Linder:2002et}
\begin{equation}
    w(z)=w_0+w_a\frac{z}{1+z},
    \label{wz}
\end{equation}
that the combination of DESI and CMB dataset shows
$w_0=-0.45\pm0.27$ and $w_a=-1.79\pm0.79$, while the constraints
on $w_0$ and $w_a$ becomes tighter when Pantheon data is
considered, which results in $w_0=0.827\pm0.063$ and
$w_a=-0.75\pm0.27$.
The relevant issues are being intensively investigated
\cite{Luongo:2024fww,Colgain:2024xqj,Cortes:2024lgw,Qu:2024lpx,Wang:2024hks,Wang:2024rjd,Carloni:2024zpl,Giare:2024smz}.

The result of DESI collaboration about an evolving DE has
astonished the scientific community, see \cite{Cortes:2024lgw} for
calling on a more comprehensive understanding of DESI result. In
view of the Hubble tension that $\Lambda$CDM model is suffering
from, we propose that it is necessary and also significant to
access the impact of DESI data on the state equation of DE in the
Hubble-tension-free cosmologies. In this paper, using recent DESI
BAO measurements combined with Planck CMB and Pantheon Plus
dataset, we find that the pre-recombination resolution of Hubble
tension would suppress the \textit{preference} of DESI for the
evolving DE, so that the CC is still $< 2\sigma$ consistent with
recent DESI data. We discuss the challenge of our results for
identifying the nature of DE.

\section{DATA AND METHODOLOGY}

Recent DESI data is based on samples of bright galaxies, LRGs,
ELGs, quasars and Ly$\alpha$ Forest at the redshift region
$0.1<z<4.2$ \cite{DESI:2024mwx,DESI:2024lzq,DESI:2024uvr}. Here,
we will use their measurements for the comoving distances
$D_M(z)/r_d$ and $D_H(z)/r_d$, where
    \begin{equation}\label{DMDH}
        D_M(z)\equiv\int_{0}^{z}{cdz'\over H(z')},\quad D_H(z)\equiv {c\over
        H(z)},
    \end{equation}
and $r_d=\int_{z_d}^{\infty}{c_s(z)\over H(z)}$ is the sound
horizon with $z_d\simeq1060$ at the baryon drag epoch and $c_s$
the speed of sound, as well as the angle-averaged quantity
$D_V/r_d$, where $D_V(z)\equiv\left(zD_M(z)^2D_H(z)\right)^{1/3}$.
    \begin{table}[htbp]
        \centering
        \begin{tabular}{c|c|cc|c}
            tracer&$z_{eff}$&$D_M/r_d$&$D_H/r_d$&$D_V/r_d$\\
            \hline
            BGS&0.30&-&-&$7.93\pm0.15$\\
            LRG&0.51&$13.62\pm0.25$&$20.98\pm0.61$&-\\
            LRG&0.71&$16.85\pm0.32$&$20.08\pm0.60$&-\\
            LRG+ELG&0.93&$21.71\pm0.28$&$17.88\pm0.35$&-\\
            ELG&1.32&$27.79\pm0.69$&$13.82\pm0.42$&-\\
            QSO&1.49&-&-&$26.07\pm0.67$\\
            Lya QSO&2.33&$39.71\pm0.94$&$8.52\pm0.17$&-\\
        \end{tabular}
        \caption{\label{DESI}Statistics for the DESI samples of the DESI DR1 BAO measurements used in this
            paper.}
    \end{table}

To study the cosmological results with DESI BAO data, we also use
\textbf{Planck 2018 CMB} (low-l and high-l TT, TE, EE spectra, and
reconstructed CMB lensing spectrum
\cite{Planck:2018vyg,Planck:2019nip,Planck:2018lbu}),
\textbf{Pantheon Plus} (consisting of 1701 light curves of 1550
spectroscopically confirmed Type Ia SN coming from 18 different
surveys \cite{Scolnic:2021amr}), and \textbf{SH0ES} ($H_0 =
73.04\pm1.04$km/s/Mpc is adopted as a Gaussian prior, equivalently
a Gaussian prior on the peak SN1a absolute magnitude
\cite{Riess:2021jrx}, see also
\cite{Camarena:2021jlr,Efstathiou:2021ocp}).



We modified the MontePython-3.6 sampler
\cite{Audren:2012wb,Brinckmann:2018cvx} and CLASS codes
\cite{Lesgourgues:2011re,Blas:2011rf} to perform our MCMC
analysis. Here, we take the axion-like EDE and AdS-EDE models as
examples of possible early resolutions of Hubble tension,
e.g.\cite{Poulin:2023lkg}. In our MCMC, the parameter sets are
\{$\omega_b$, $\omega_{cdm}$, $H_0$, $\ln10^{10}A_s$, $n_s$,
$\tau_{reio}$, $w_0$, $w_a$, $\log_{10}a_c$, $f_\mathrm{EDE}$,
$\theta_i$\} for axion-like EDE \cite{Poulin:2018cxd}, where
$\theta_i\equiv\phi_i/f$ and $f_\mathrm{EDE}$ is the energy
fraction of EDE at $z_c$, and \{$\omega_b$, $\omega_{cdm}$, $H_0$,
$\ln10^{10}A_s$, $n_s$, $\tau_{reio}$, $w_0$, $w_a$, $\ln(1+z_c)$,
$f_\mathrm{EDE}$\} for AdS-EDE (see
Ref.\cite{Ye:2020btb,Ye:2020oix,Jiang:2021bab}), respectively. In
Table.\ref{prior} we list the flat priors used. We adopt a
Gelman-Rubin convergence criterion with a threshold $R-1<0.05$.

    \begin{table}[htbp]
    \centering
    \begin{tabular}{cc}
        \hline
        Parameters&Prior\\
        \hline
        $100\omega_b$&[None, None]\\
        $\omega_{cdm}$&[None, None]\\
        $H_0$&[65, 80]\\
        $\ln10^{10}A_s$&[None, None]\\
        $n_s$&[None,None]\\
        $\tau_{reio}$&[0.004, None]\\
        \hline
        $f_\mathrm{EDE}$&[$10^{-4}$, 0.3]\\
        $w_0$&[-2, 0.34]\\
        $w_a$&[-3,2]\\
        \hline
    \end{tabular}
    \caption{\label{prior} The priors of parameters we adopt in MCMC analysis.}
 \end{table}

\section{RESULTS}
    \begin{table*}[htbp]
    \centering
    \begin{tabular}{c|c|c|c}
        \hline
        \multirow{2}{*}{Parameters}&\multirow{2}{*}{$w_0w_a$CDM}& $w_0w_a$CDM & $w_0w_a$CDM\\
        &&+axion-like EDE&+AdS EDE\\
        \hline
        $100\omega_b$&2.234(2.236)$\pm$0.015&2.279(2.304)$\pm$0.025&2.332(2.348)$\pm$0.020\\
        $\omega_{cdm}$&0.120(0.120)$\pm$0.001&0.130(0.132)$\pm$0.003&0.135(0.132)$\pm$0.002\\
        $H_0$&70.03(69.45)$\pm$0.61&71.90(72.31)$\pm$0.78&72.71(72.83)$\pm$0.68\\
        $\ln10^{10}A_s$&3.035(3.044)$\pm$0.015&3.053(3.049)$\pm$0.015&3.072(3.079)$\pm$0.015\\
        $n_s$&0.965(0.965)$\pm$0.004&0.987(0.996)$\pm$0.008&0.995(0.999)$\pm$0.008\\
        $\tau_{reio}$&0.052(0.056)$\pm$0.007&0.054(0.052)$\pm$0.008&0.054(0.054)$\pm$0.008\\
        \hline
        $w_0$&-0.831(-0.781)$\pm$0.068&-0.852(-0.811)$\pm$0.064&-0.861(-0.861)$\pm$0.056\\
        $w_a$&-0.953(-1.072)$\pm$0.308&-0.636(-0.684)$\pm$0.275&-0.569(-0.566)$\pm$0.228\\
        \hline
        $\Omega_m$&0.290(0.295)$\pm$0.006&0.296(0.297)$\pm$0.006&0.300(0.297)$\pm$0.005\\
        $S_8$&0.829(0.836)$\pm$0.011&0.847(0.843)$\pm$0.012&0.863(0.855)$\pm$0.012\\
        \hline
        $\chi^2_\mathrm{CMB}$&2769.68&2771.05&2777.17\\
        $\chi^2_\mathrm{DESI}$&19.15&16.49&15.67\\
        $\chi^2_{\mathrm{Pantheon+}H_0}$&1306.38&1289.45&1288.12\\
        \hline
        $\chi^2_\mathrm{tot}$&4095.21&4076.99&4080.98\\
        $\Delta\chi^2_\mathrm{tot}$&0&-18.22&-14.23\\
        \hline
    \end{tabular}
\caption{\label{EDE} Mean (bestfit) values and 1$\sigma$ regions
of the parameters of the $w_0w_a$CDM model with axion-like EDE and
AdS-EDE, respectively, fitting to Planck18+DESI+Pantheon+SH0ES
dataset.}
    \end{table*}

In Table.\ref{EDE}, we present our MCMC results. The $1\sigma$ and
$2\sigma$ marginalized posterior distributions of the
corresponding parameters are showed in Appendix-\ref{appendix}.
In both $w_0w_a$CDM+EDE models, the bestfit values of $H_0$ are
$H_0\sim72.3$ and $72.8$km/s/Mpc, respectively, both are $1\sigma$
consistent with recent local measurement of $H_0$
\cite{Breuval:2024lsv}, and slightly higher than those with
pre-DESI data, see e.g.\cite{Wang:2022jpo}. In
$w_0w_a$CDM+(axion-like)EDE model, $S_8$ is smaller than that with
pre-DESI data. Actually DESI itself prefers a lower $S_8$, see
e.g.\cite{Qu:2024lpx}. However, in the $w_0w_a$CDM+(AdS-)EDE model
the exacerbation of $S_8$ caused by a larger $H_0$ offsets the
effect of DESI, with a $S_8$ similar to that with pre-DESI data.



\begin{figure*}
    \includegraphics[width=1.8\columnwidth]{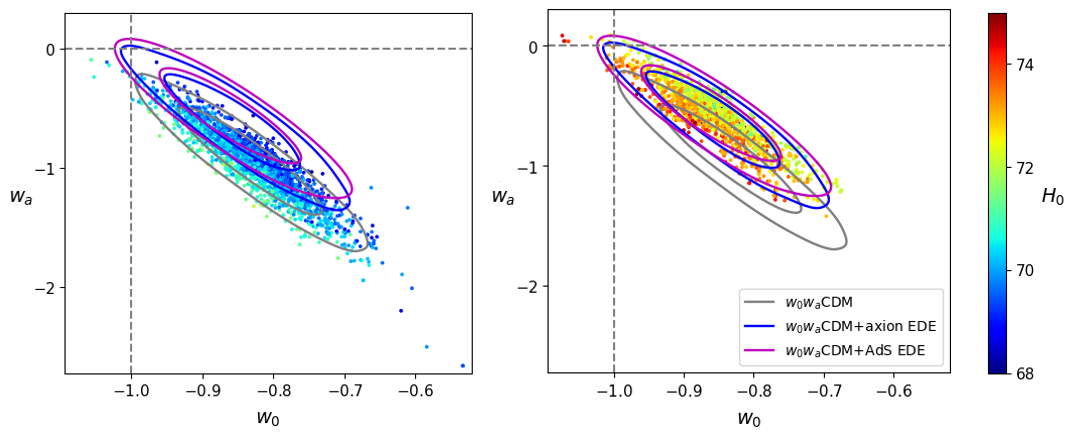}
\caption{2D contours at 68\% and 95\% CL of the parameters
$w_0$-$w_a$ for the $w_0w_a$CDM model with axion-like EDE and
AdS-EDE, respectively, fitting to Planck18+DESI+Pantheon+SH0ES
dataset. Left panel: scattered plot for the $w_0w_a$CDM model.
Right panel: scattered plot for the $w_0w_a$CDM+(AdS-)EDE model.}
\label{w0waH0}
\end{figure*}

It is significant to check the results about DE. In the
$w_0w_a$CDM+EDE model, $w_0>-1$ and $w_a<0$ are still preferred,
but both are closer to $w_0=-1$ and $w_a=0$, respectively, in
particular the CC is now $<2\sigma$ consistent. In
Fig.\ref{w0waH0}, we show the $1\sigma$ and $2\sigma$ marginalized
posterior distributions of $w_0-w_a$ with the colorbar of $H_0$.
It seems that a higher $H_0$ is pushing $w_0$ and $w_a$ closer to
$w_0=-1$ and $w_a=0$. This could be partly explained as follows.
According to the parameterization (\ref{wz}), we have
\begin{widetext}
\begin{equation}
H(z)/H_0=\left[\Omega_m(1+z)^3+(1-\Omega_m)(1+z)^{3(1+w_0+w_a)}e^{-3w_az/(1+z)}\right]^{1/2}.
\label{HzH0}
\end{equation}
\end{widetext}
In light of Eqs.(\ref{DMDH}) and (\ref{HzH0}), we see that
$D_M/r_d$ and $D_H/r_d$) are dependent on $H_0$ and $w_0$ ($w_a$).
In certain sense, when $D_M/r_d$ ($D_H/r_d$) is fixed, a high
$H_0$ will help to suppress the requirements of data for $w_0>-1$
and $w_a<0$, and make both closer to $w_0=-1$ and $w_a=0$.

\begin{figure*}
    \includegraphics[width=1.8\columnwidth]{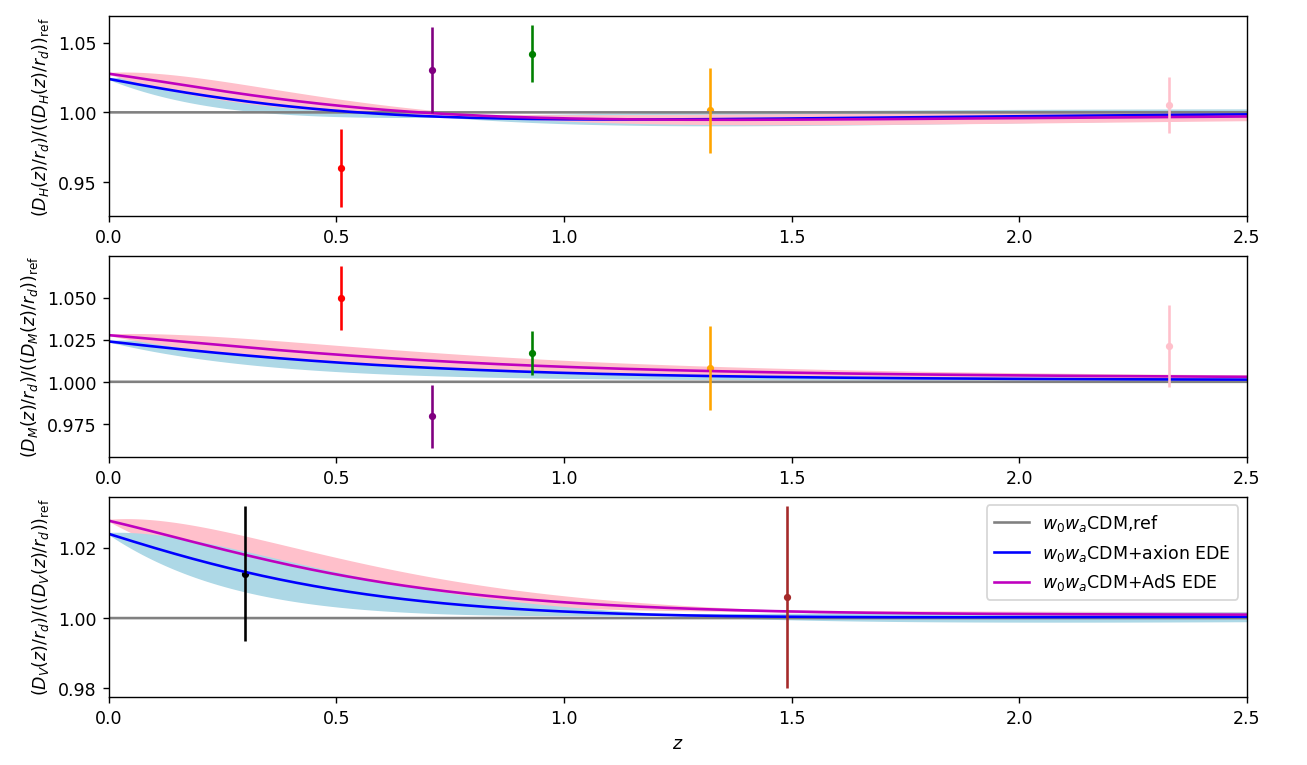}
\caption{Residuals of $D_H(z)/r_d$, $D_M(z)/r_d$ and $D_V(z)/r_d$
for the bestfit values of the $w_0w_a$CDM+EDE model with respect
to the $w_0w_a$CDM model fitting to
Planck18+DESI+PantheonPlus+SH0ES datasets. The blue and pink
shadows are $1\sigma$ regions of $D_H(z)/r_d$, $D_M(z)/r_d$ and
$D_V(z)/r_d$, respectively. The DESI data points and error bars
are as listed in Table.\ref{DESI}. } \label{DHDV}
\end{figure*}

It is also noted that though in the $w_0w_a$CDM+EDE model the fit
to CMB is slightly exacerbated, unexpectedly the fit to DESI BAO
is improved, with $\Delta\chi^2_\mathrm{DESI}\thickapprox -2.7$
and $-3.5$ for axion-like EDE and AdS-EDE, respectively. This in
certain sense might imply a slightly support of the DESI data
itself for the pre-recombination resolutions of Hubble tension,
see also Fig.\ref{DHDV}.

\section{Discussion}

In summary, we accessed the impact of DESI data on the state
equation of DE in the Hubble-tension-free cosmologies. We found
that though $w_0>-1$ and $w_a<0$ are still preferred, the
pre-recombination resolution of Hubble tension suppress the
\textit{preference} of DESI for the evolving DE, so that the CC is
$< 2\sigma$ consistent. This suggests that the evidence against CC
is likely to disappear when certain resolution of Hubble tension
is adopted.

The concordance $\Lambda$CDM model is the simplest description of
our Universe, which can naturally explain most of cosmological and
astrophysical observations. Here, with recent DESI data we found
the possibility that DE is still CC, which suggests that the
$\gtrsim 3\sigma$ evidence for evolving DE might be an artefact
without considering the attack of the Hubble tension. It is also
interesting to explore the effects of different resolutions of
Hubble tension (see
e.g.\cite{Knox:2019rjx,Perivolaropoulos:2021jda,DiValentino:2021izs,Vagnozzi:2023nrq})
on the results of DESI about DE, e.g.recent \cite{Giare:2024smz}.
According to our result, not considering the Hubble tensions which
the $\Lambda$CDM (so $w_0w_a$CDM) model is suffering from might
bias our understanding for the nature of DE, thus the claim of
ruling out the CC needs to be more cautious regarding not only the
recent data but also the cosmological tensions.

The results of DESI about DE would also affect our understanding
for the very early universe. It is well-known that inflation is
the current paradigm of early universe, which predicts (nearly)
scale-invariant scalar perturbation. Though based on the
$\Lambda$CDM model the Planck collaboration showed the scalar
spectral index is $n_s\approx 0.965$ \cite{Planck:2018jri},
$n_s=1$ ($n_s-1\sim {\cal O} (0.001)$) is favored
\cite{Ye:2020btb,Ye:2021nej,Jiang:2022uyg,Smith:2022hwi,Jiang:2022qlj,Peng:2023bik}
when the EDE resolution of Hubble tension is considered, see also
\cite{DiValentino:2018zjj,Giare:2022rvg}, and see recent
Refs.\cite{Kallosh:2022ggf,Braglia:2020bym,Ye:2022efx,Jiang:2023bsz,Braglia:2022phb,DAmico:2021fhz,
Takahashi:2021bti,Giare:2023wzl,Fu:2023tfo,Giare:2024akf} for its
implications for inflation models and discussions. Here, with DESI
data we find that the shift towards $n_s=1$ persists, see
Table.\ref{EDE}. It will be interesting to revisit inflation
models in the light of new constraints on $n_s$.

Currently, though it is still too early to say whether DE is
evolving or not and whether $n_s=1$ or not, hopefully the further
data of DESI and upcoming Euclid \cite{Euclid:2024yrr} combined
with CMB and SN dataset might tell us the corresponding answers.

\section*{Acknowledgments}

We acknowledge the use of publicly available codes AxiCLASS
(\url{https://github.com/PoulinV/AxiCLASS}) and classmultiscf
(\url{https://github.com/genye00/class_multiscf.git}). We thank
Bin Hu for valuable discussion. YSP is supported by NSFC,
No.12075246, National Key Research and Development Program of
China, No. 2021YFC2203004, and the Fundamental Research Funds for
the Central Universities.

\appendix

\section{Results of the $w_0w_a$CDM model w/o SH0ES}

\begin{figure*}
    \includegraphics[width=1.2\columnwidth]{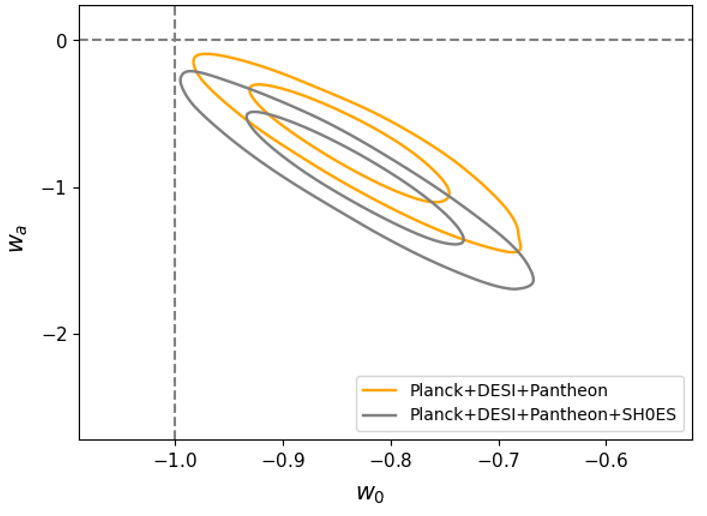}
\caption{2D contours at 68\% and 95\% CL of the parameters
$w_0$-$w_a$ for the $w_0w_a$CDM model fitting to
Planck18+DESI+Pantheon dataset and Planck18+DESI+Pantheon+SH0ES
dataset, respectively.} \label{w0wawithoutH0}
\end{figure*}

In order to check the consistence of our modified likelihood, we
make a test for the $w_0w_a$CDM model with only
Planck+DESI+PantheonPlus dataset. Mean (bestfit) values and
1$\sigma$ regions of the parameters of $w_0w_a$CDM model are
listed in Table.\ref{withoutH0}. The $1\sigma$ and $2\sigma$
marginalized posterior distributions of $w_0-w_a$ is showed in
Fig.\ref{w0wawithoutH0}. It can be seen that the results without
SH0ES are completely consistent with those in DESI paper
\cite{DESI:2024mwx}.

    \begin{table*}[htbp]
    \centering
    \begin{tabular}{c|c|c}
        \hline
        \multicolumn{3}{c}{$w_0w_a$CDM}\\
        \hline
        Parameters&Planck+DESI+Pantheon&Planck+DESI+Pantheon+SH0ES\\
        \hline
        $100\omega_b$&2.222(2.205)$\pm$0.013&2.234(2.236)$\pm$0.015\\
        $\omega_{cdm}$&0.119(0.119)$\pm$0.001&0.120(0.120)$\pm$0.001\\
        $H_0$&67.86(67.79)$\pm$0.73&70.03(69.45)$\pm$0.61\\
        $\ln10^{10}A_s$&3.036(3.037)$\pm$0.014&3.035(3.044)$\pm$0.015\\
        $n_s$&0.964(0.961)$\pm$0.004&0.965(0.965)$\pm$0.004\\
        $\tau_{reio}$&0.052(0.053)$\pm$0.007&0.052(0.056)$\pm$0.007\\
        \hline
        $w_0$&-0.844(-0.855)$\pm$0.071&-0.831(-0.781)$\pm$0.068\\
        $w_a$&-0.670(-0.640)$\pm$0.306&-0.953(-1.072)$\pm$0.308\\
        \hline
        $\Omega_m$&0.309(0.309)$\pm$0.007&0.290(0.295)$\pm$0.006\\
        $S_8$&0.828(0.832)$\pm$0.015&0.829(0.836)$\pm$0.011\\
        \hline
    \end{tabular}
    \caption{ \label{withoutH0} Mean (bestfit) values and 1$\sigma$ regions
of the parameters of the $w_0w_a$CDM model fitting to
Planck18+DESI+Pantheon dataset and Planck18+DESI+Pantheon+SH0ES
dataset, respectively.}
\end{table*}

\section{2D contour results of $w_0w_a$CDM+EDE models}\label{appendix}

\begin{figure*}
    \includegraphics[width=1.8\columnwidth]{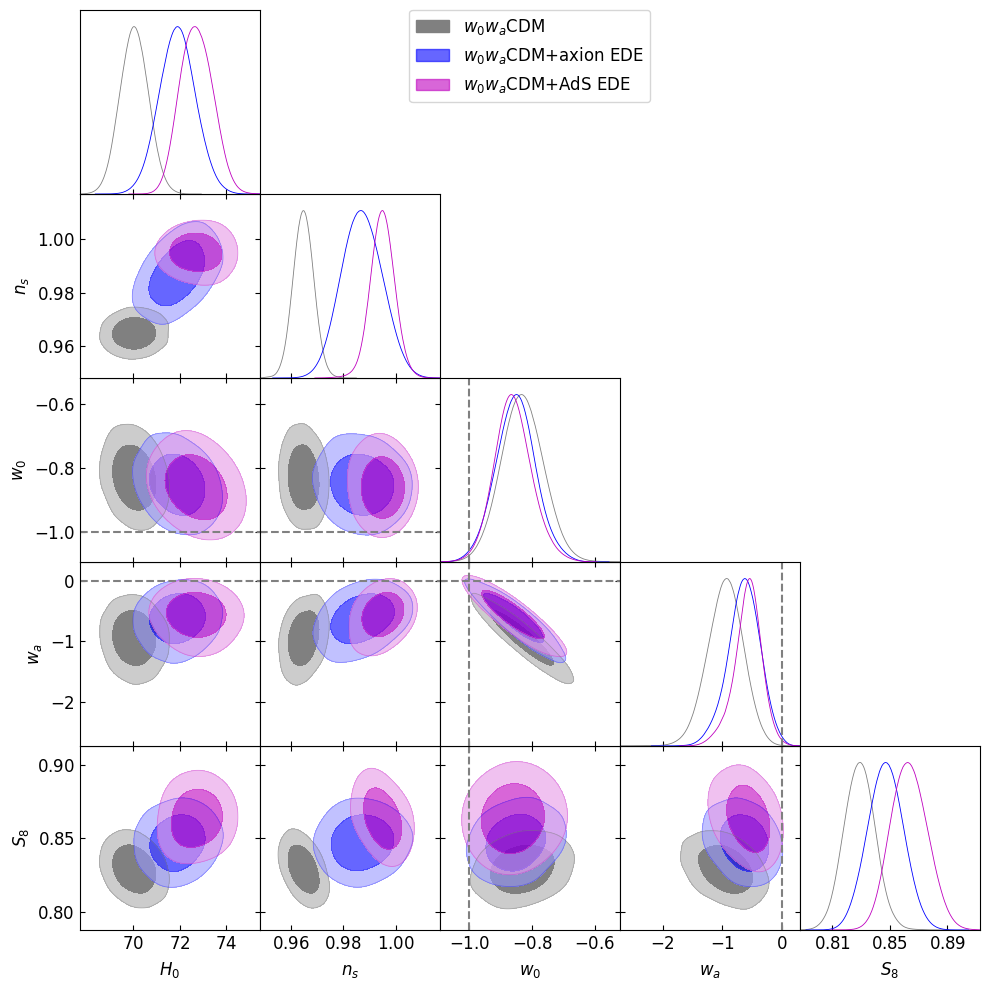}
\caption{\label{MC}2D contours at 68\% and 95\% CL for the
parameters of the $w_0w_a$CDM model with axion-like EDE and
AdS-EDE, respectively, fitting to Planck18+DESI+Pantheon+SH0ES
dataset.} \label{MCMCresult}
\end{figure*}

In Fig.\ref{MCMCresult}, we present the $1\sigma$ and $2\sigma$
marginalized posterior distributions of main parameters for the
$w_0w_a$CDM model with axion-like EDE and AdS-EDE, respectively,
fitting to Planck18+DESI+Pantheon+SH0ES dataset.

\begin{thebibliography}{99}



        \bibitem{Riess:2019qba}
        A.~G.~Riess,
        Nature Rev. Phys. \textbf{2}, no.1, 10-12 (2019)
        doi:10.1038/s42254-019-0137-0
        [arXiv:2001.03624 [astro-ph.CO]].


\bibitem{Verde:2019ivm}
L.~Verde, T.~Treu and A.~G.~Riess,
Nature Astron. \textbf{3}, 891
doi:10.1038/s41550-019-0902-0
[arXiv:1907.10625 [astro-ph.CO]].

        \bibitem{Breuval:2024lsv}
        L.~Breuval, A.~G.~Riess, S.~Casertano, W.~Yuan, L.~M.~Macri, M.~Romaniello, Y.~S.~Murakami, D.~Scolnic, G.~S.~Anand and I.~Soszy\'nski,
        [arXiv:2404.08038 [astro-ph.CO]].

        \bibitem{Planck:2018vyg}
        N.~Aghanim \textit{et al.} [Planck],
        Astron. Astrophys. \textbf{641}, A6 (2020)
        [erratum: Astron. Astrophys. \textbf{652}, C4 (2021)]
        doi:10.1051/0004-6361/201833910
        [arXiv:1807.06209 [astro-ph.CO]].




\bibitem{Knox:2019rjx}
L.~Knox and M.~Millea,
Phys. Rev. D \textbf{101}, no.4, 043533 (2020)
doi:10.1103/PhysRevD.101.043533
[arXiv:1908.03663 [astro-ph.CO]].


\bibitem{DiValentino:2021izs}
E.~Di Valentino, O.~Mena, S.~Pan, L.~Visinelli, W.~Yang, A.~Melchiorri, D.~F.~Mota, A.~G.~Riess and J.~Silk,
Class. Quant. Grav. \textbf{38}, no.15, 153001 (2021)
doi:10.1088/1361-6382/ac086d
[arXiv:2103.01183 [astro-ph.CO]].

\bibitem{Perivolaropoulos:2021jda}
L.~Perivolaropoulos and F.~Skara,
New Astron. Rev. \textbf{95}, 101659 (2022)
doi:10.1016/j.newar.2022.101659
[arXiv:2105.05208 [astro-ph.CO]].

\bibitem{Vagnozzi:2023nrq}
S.~Vagnozzi,
Universe \textbf{9} (2023) no.9, 393 doi:10.3390/universe9090393
[arXiv:2308.16628 [astro-ph.CO]].


        \bibitem{Karwal:2016vyq}
        T.~Karwal and M.~Kamionkowski,
        Phys. Rev. D \textbf{94}, no.10, 103523 (2016)
        doi:10.1103/PhysRevD.94.103523 [arXiv:1608.01309 [astro-ph.CO]].

        \bibitem{Poulin:2018cxd}
        V.~Poulin, T.~L.~Smith, T.~Karwal and M.~Kamionkowski,
        Phys. Rev. Lett. \textbf{122}, no.22, 221301 (2019)
        doi:10.1103/PhysRevLett.122.221301
        [arXiv:1811.04083 [astro-ph.CO]].

        \bibitem{Smith:2019ihp}
        T.~L.~Smith, V.~Poulin and M.~A.~Amin,
        Phys. Rev. D \textbf{101}, no.6, 063523 (2020)
        doi:10.1103/PhysRevD.101.063523
        [arXiv:1908.06995 [astro-ph.CO]].


        \bibitem{Ye:2020btb}
        G.~Ye and Y.~S.~Piao,
        Phys. Rev. D \textbf{101}, no.8, 083507 (2020)
        doi:10.1103/PhysRevD.101.083507
        [arXiv:2001.02451 [astro-ph.CO]].

        \bibitem{Ye:2020oix}
        G.~Ye and Y.~S.~Piao,
        Phys. Rev. D \textbf{102}, no.8, 083523 (2020)
        doi:10.1103/PhysRevD.102.083523 [arXiv:2008.10832 [astro-ph.CO]].


        \bibitem{DAmico:2020ods}
        G.~D'Amico, L.~Senatore, P.~Zhang and H.~Zheng,
        JCAP \textbf{05}, 072 (2021)
        doi:10.1088/1475-7516/2021/05/072
        [arXiv:2006.12420 [astro-ph.CO]].

        \bibitem{Krishnan:2020obg}
        C.~Krishnan, E.~\'O.~Colg\'ain, Ruchika, A.~A.~Sen, M.~M.~Sheikh-Jabbari and T.~Yang,
        Phys. Rev. D \textbf{102}, no.10, 103525 (2020)
        doi:10.1103/PhysRevD.102.103525
        [arXiv:2002.06044 [astro-ph.CO]].

        \bibitem{Nunes:2021ipq}
        R.~C.~Nunes and S.~Vagnozzi,
        Mon. Not. Roy. Astron. Soc. \textbf{505}, no.4, 5427-5437 (2021)
        doi:10.1093/mnras/stab1613 [arXiv:2106.01208 [astro-ph.CO]].

        \bibitem{Hill:2020osr}
        J.~C.~Hill, E.~McDonough, M.~W.~Toomey and S.~Alexander,
        Phys. Rev. D \textbf{102}, no.4, 043507 (2020)
        doi:10.1103/PhysRevD.102.043507
        [arXiv:2003.07355 [astro-ph.CO]].

        \bibitem{Ivanov:2020ril}
        M.~M.~Ivanov, E.~McDonough, J.~C.~Hill, M.~Simonovi\'c, M.~W.~Toomey, S.~Alexander and M.~Zaldarriaga,
        Phys. Rev. D \textbf{102}, no.10, 103502 (2020)
        doi:10.1103/PhysRevD.102.103502
        [arXiv:2006.11235 [astro-ph.CO]].

        \bibitem{Poulin:2022sgp}
        V.~Poulin, J.~L.~Bernal, E.~Kovetz and M.~Kamionkowski,
        [arXiv:2209.06217 [astro-ph.CO]].

        \bibitem{Allali:2021azp}
        I.~J.~Allali, M.~P.~Hertzberg and F.~Rompineve,
        Phys. Rev. D \textbf{104}, no.8, L081303 (2021)
        doi:10.1103/PhysRevD.104.L081303 [arXiv:2104.12798 [astro-ph.CO]].

        \bibitem{Ye:2021iwa}
        G.~Ye, J.~Zhang and Y.~S.~Piao,
        [arXiv:2107.13391 [astro-ph.CO]].

        \bibitem{Alexander:2022own}
        S.~Alexander, H.~Bernardo and M.~W.~Toomey,
        [arXiv:2207.13086 [astro-ph.CO]].

        \bibitem{FrancoAbellan:2021sxk}
        G.~Franco Abell\'an, R.~Murgia and V.~Poulin,
        Phys. Rev. D \textbf{104}, no.12, 12 (2021)
        doi:10.1103/PhysRevD.104.123533 [arXiv:2102.12498 [astro-ph.CO]].

        \bibitem{Clark:2021hlo}
        S.~J.~Clark, K.~Vattis, J.~Fan and S.~M.~Koushiappas,
        [arXiv:2110.09562 [astro-ph.CO]].

        \bibitem{Simon:2022ftd}
        T.~Simon, G.~Franco Abell\'an, P.~Du, V.~Poulin and Y.~Tsai,
        Phys. Rev. D \textbf{106}, no.2, 023516 (2022)
        doi:10.1103/PhysRevD.106.023516 [arXiv:2203.07440 [astro-ph.CO]].

        \bibitem{Reeves:2022aoi}
        A.~Reeves, L.~Herold, S.~Vagnozzi, B.~D.~Sherwin and E.~G.~M.~Ferreira,
        Mon. Not. Roy. Astron. Soc. \textbf{520}, no.3, 3688-3695 (2023)
        doi:10.1093/mnras/stad317
        [arXiv:2207.01501 [astro-ph.CO]].

\bibitem{Wang:2022bmk}
H.~Wang and Y.~S.~Piao,
Phys. Rev. D \textbf{108} (2023) no.8, 8
doi:10.1103/PhysRevD.108.083516 [arXiv:2209.09685 [astro-ph.CO]].

        \bibitem{Chacko:2016kgg}
        Z.~Chacko, Y.~Cui, S.~Hong, T.~Okui and Y.~Tsai,
        JHEP \textbf{12}, 108 (2016) doi:10.1007/JHEP12(2016)108
        [arXiv:1609.03569 [astro-ph.CO]].

        \bibitem{Buen-Abad:2022kgf}
        M.~A.~Buen-Abad, Z.~Chacko, C.~Kilic, G.~Marques-Tavares and
        T.~Youn,
        [arXiv:2208.05984 [hep-ph]].

\bibitem{Teng:2021cvy}
Y.~P.~Teng, W.~Lee and K.~W.~Ng,
Phys. Rev. D \textbf{104} (2021) no.8, 083519
doi:10.1103/PhysRevD.104.083519 [arXiv:2105.02667 [astro-ph.CO]].

\bibitem{Dainotti:2021pqg}
M.~G.~Dainotti, B.~De Simone, T.~Schiavone, G.~Montani, E.~Rinaldi
and G.~Lambiase,
Astrophys. J. \textbf{912} (2021) no.2, 150
doi:10.3847/1538-4357/abeb73 [arXiv:2103.02117 [astro-ph.CO]].

\bibitem{Dainotti:2022bzg}
M.~G.~Dainotti, B.~De Simone, T.~Schiavone, G.~Montani,
E.~Rinaldi, G.~Lambiase, M.~Bogdan and S.~Ugale,
Galaxies \textbf{10} (2022) no.1, 24 doi:10.3390/galaxies10010024
[arXiv:2201.09848 [astro-ph.CO]].

\bibitem{Montani:2024pou}
G.~Montani and N.~Carlevaro,
[arXiv:2404.15977 [gr-qc]].

\bibitem{Frieman:2008sn}
J.~Frieman, M.~Turner and D.~Huterer,
Ann. Rev. Astron. Astrophys. \textbf{46} (2008), 385-432
doi:10.1146/annurev.astro.46.060407.145243 [arXiv:0803.0982
[astro-ph]].

\bibitem{Nojiri:2017ncd}
S.~Nojiri, S.~D.~Odintsov and V.~K.~Oikonomou,
Phys. Rept. \textbf{692} (2017), 1-104
doi:10.1016/j.physrep.2017.06.001 [arXiv:1705.11098 [gr-qc]].

\bibitem{Carroll:2000fy}
S.~M.~Carroll,
Living Rev. Rel. \textbf{4} (2001), 1 doi:10.12942/lrr-2001-1
[arXiv:astro-ph/0004075 [astro-ph]].

\bibitem{Huterer:2017buf}
D.~Huterer and D.~L.~Shafer,
Rept. Prog. Phys. \textbf{81} (2018) no.1, 016901
doi:10.1088/1361-6633/aa997e [arXiv:1709.01091 [astro-ph.CO]].


        \bibitem{DESI:2024mwx}
        A.~G.~Adame \textit{et al.} [DESI],
        [arXiv:2404.03002 [astro-ph.CO]].

        \bibitem{DESI:2024uvr}
        A.~G.~Adame \textit{et al.} [DESI],
        [arXiv:2404.03000 [astro-ph.CO]].

        \bibitem{DESI:2024lzq}
        A.~G.~Adame \textit{et al.} [DESI],
        [arXiv:2404.03001 [astro-ph.CO]].

        \bibitem{Chevallier:2000qy}
        M.~Chevallier and D.~Polarski,
        Int. J. Mod. Phys. D \textbf{10} (2001), 213-224
        doi:10.1142/S0218271801000822
        [arXiv:gr-qc/0009008 [gr-qc]].

        \bibitem{Linder:2002et}
        E.~V.~Linder,
        Phys. Rev. Lett. \textbf{90} (2003), 091301
        doi:10.1103/PhysRevLett.90.091301
        [arXiv:astro-ph/0208512 [astro-ph]].

\bibitem{Cortes:2024lgw}
M.~Cort\^es and A.~R.~Liddle,
[arXiv:2404.08056 [astro-ph.CO]].

\bibitem{Luongo:2024fww}
O.~Luongo and M.~Muccino,
[arXiv:2404.07070 [astro-ph.CO]].



        \bibitem{Qu:2024lpx}
        F.~J.~Qu, K.~M.~Surrao, B.~Bolliet, J.~C.~Hill, B.~D.~Sherwin and H.~T.~Jense,
        [arXiv:2404.16805 [astro-ph.CO]].


        \bibitem{Carloni:2024zpl}
        Y.~Carloni, O.~Luongo and M.~Muccino,
        [arXiv:2404.12068 [astro-ph.CO]].

\bibitem{Wang:2024hks}
D.~Wang,
[arXiv:2404.06796 [astro-ph.CO]].

        \bibitem{Wang:2024rjd}
        D.~Wang,
        [arXiv:2404.13833 [astro-ph.CO]].

        \bibitem{Colgain:2024xqj}
        E.~\'O.~Colg\'ain, M.~G.~Dainotti, S.~Capozziello, S.~Pourojaghi, M.~M.~Sheikh-Jabbari and D.~Stojkovic,
        [arXiv:2404.08633 [astro-ph.CO]].

\bibitem{Giare:2024smz}
W.~Giar\`e, M.~A.~Sabogal, R.~C.~Nunes and E.~Di Valentino,
[arXiv:2404.15232 [astro-ph.CO]].



        \bibitem{Planck:2019nip}
        N.~Aghanim \textit{et al.} [Planck],
        Astron. Astrophys. \textbf{641} (2020), A5
        doi:10.1051/0004-6361/201936386
        [arXiv:1907.12875 [astro-ph.CO]].

        \bibitem{Planck:2018lbu}
        N.~Aghanim \textit{et al.} [Planck],
        Astron. Astrophys. \textbf{641} (2020), A8
        doi:10.1051/0004-6361/201833886
        [arXiv:1807.06210 [astro-ph.CO]].

        \bibitem{Scolnic:2021amr}
        D.~Scolnic, D.~Brout, A.~Carr, A.~G.~Riess, T.~M.~Davis, A.~Dwomoh, D.~O.~Jones, N.~Ali, P.~Charvu and R.~Chen, \textit{et al.}
        Astrophys. J. \textbf{938} (2022) no.2, 113
        doi:10.3847/1538-4357/ac8b7a
        [arXiv:2112.03863 [astro-ph.CO]].

        \bibitem{Riess:2021jrx}
        A.~G.~Riess, W.~Yuan, L.~M.~Macri, D.~Scolnic, D.~Brout, S.~Casertano, D.~O.~Jones, Y.~Murakami, L.~Breuval and T.~G.~Brink, \textit{et al.}
        Astrophys. J. Lett. \textbf{934}, no.1, L7 (2022)
        doi:10.3847/2041-8213/ac5c5b
        [arXiv:2112.04510 [astro-ph.CO]].


    \bibitem{Camarena:2021jlr}
    D.~Camarena and V.~Marra,
    Mon. Not. Roy. Astron. Soc. \textbf{504} (2021), 5164-5171
    doi:10.1093/mnras/stab1200
    [arXiv:2101.08641 [astro-ph.CO]].

    \bibitem{Efstathiou:2021ocp}
    G.~Efstathiou,
    Mon. Not. Roy. Astron. Soc. \textbf{505} (2021) no.3, 3866-3872
    doi:10.1093/mnras/stab1588
    [arXiv:2103.08723 [astro-ph.CO]].

\bibitem{Jiang:2021bab}
J.~Q.~Jiang and Y.~S.~Piao,
Phys. Rev. D \textbf{104} (2021) no.10, 103524
doi:10.1103/PhysRevD.104.103524 [arXiv:2107.07128 [astro-ph.CO]].

        \bibitem{Wang:2022jpo}
        H.~Wang and Y.~S.~Piao,
        Phys. Lett. B \textbf{832} (2022), 137244
        doi:10.1016/j.physletb.2022.137244
        [arXiv:2201.07079 [astro-ph.CO]].







        \bibitem{Audren:2012wb}
        B.~Audren, J.~Lesgourgues, K.~Benabed and S.~Prunet,
        JCAP \textbf{02} (2013), 001
        doi:10.1088/1475-7516/2013/02/001
        [arXiv:1210.7183 [astro-ph.CO]].

        \bibitem{Brinckmann:2018cvx}
        T.~Brinckmann and J.~Lesgourgues,
        Phys. Dark Univ. \textbf{24} (2019), 100260
        doi:10.1016/j.dark.2018.100260
        [arXiv:1804.07261 [astro-ph.CO]].

        \bibitem{Lesgourgues:2011re}
        J.~Lesgourgues,
        [arXiv:1104.2932 [astro-ph.IM]].

        \bibitem{Blas:2011rf}
        D.~Blas, J.~Lesgourgues and T.~Tram,
        JCAP \textbf{07} (2011), 034
        doi:10.1088/1475-7516/2011/07/034
        [arXiv:1104.2933 [astro-ph.CO]].

\bibitem{Poulin:2023lkg}
V.~Poulin, T.~L.~Smith and T.~Karwal,
Phys. Dark Univ. \textbf{42} (2023), 101348
doi:10.1016/j.dark.2023.101348 [arXiv:2302.09032 [astro-ph.CO]].


\bibitem{Planck:2018jri}
Y.~Akrami \textit{et al.} [Planck],
Astron. Astrophys. \textbf{641} (2020), A10
doi:10.1051/0004-6361/201833887 [arXiv:1807.06211 [astro-ph.CO]].


\bibitem{Ye:2021nej}
G.~Ye, B.~Hu and Y.~S.~Piao,
Phys. Rev. D \textbf{104} (2021) no.6, 063510
doi:10.1103/PhysRevD.104.063510 [arXiv:2103.09729 [astro-ph.CO]].

\bibitem{Jiang:2022uyg}
J.~Q.~Jiang and Y.~S.~Piao,
Phys. Rev. D \textbf{105} (2022) no.10, 103514
doi:10.1103/PhysRevD.105.103514 [arXiv:2202.13379 [astro-ph.CO]].

\bibitem{Jiang:2022qlj}
J.~Q.~Jiang, G.~Ye and Y.~S.~Piao,
[arXiv:2210.06125 [astro-ph.CO]].

\bibitem{Smith:2022hwi}
T.~L.~Smith, M.~Lucca, V.~Poulin, G.~F.~Abellan, L.~Balkenhol,
K.~Benabed, S.~Galli and R.~Murgia,
Phys. Rev. D \textbf{106} (2022) no.4, 043526
doi:10.1103/PhysRevD.106.043526 [arXiv:2202.09379 [astro-ph.CO]].

\bibitem{Peng:2023bik}
Z.~Y.~Peng and Y.~S.~Piao,
[arXiv:2308.01012 [astro-ph.CO]].

\bibitem{Kallosh:2022ggf}
R.~Kallosh and A.~Linde,
Phys. Rev. D \textbf{106} (2022) no.2, 023522
doi:10.1103/PhysRevD.106.023522 [arXiv:2204.02425 [hep-th]].

\bibitem{Braglia:2020bym}
M.~Braglia, W.~T.~Emond, F.~Finelli, A.~E.~Gumrukcuoglu and
K.~Koyama,
Phys. Rev. D \textbf{102} (2020) no.8, 083513
doi:10.1103/PhysRevD.102.083513 [arXiv:2005.14053 [astro-ph.CO]].


\bibitem{Ye:2022efx}
G.~Ye, J.~Q.~Jiang and Y.~S.~Piao,
Phys. Rev. D \textbf{106} (2022) no.10, 103528
doi:10.1103/PhysRevD.106.103528 [arXiv:2205.02478 [astro-ph.CO]].

\bibitem{Jiang:2023bsz}
J.~Q.~Jiang, G.~Ye and Y.~S.~Piao,
[arXiv:2303.12345 [astro-ph.CO]].

\bibitem{Braglia:2022phb}
M.~Braglia, A.~Linde, R.~Kallosh and F.~Finelli,
JCAP \textbf{04} (2023), 033 doi:10.1088/1475-7516/2023/04/033
[arXiv:2211.14262 [astro-ph.CO]].


\bibitem{DAmico:2021fhz}
G.~D'Amico, N.~Kaloper and A.~Westphal,
Phys. Rev. D \textbf{105}, no.10, 103527 (2022)
doi:10.1103/PhysRevD.105.103527 [arXiv:2112.13861 [hep-th]].

\bibitem{Takahashi:2021bti}
F.~Takahashi and W.~Yin,
Phys. Lett. B \textbf{830}, 137143 (2022)
doi:10.1016/j.physletb.2022.137143 [arXiv:2112.06710
[astro-ph.CO]].

\bibitem{Giare:2023wzl}
W.~Giar\`e, S.~Pan, E.~Di Valentino, W.~Yang, J.~de Haro and
A.~Melchiorri,
JCAP \textbf{09} (2023), 019 doi:10.1088/1475-7516/2023/09/019
[arXiv:2305.15378 [astro-ph.CO]].

\bibitem{Fu:2023tfo}
C.~Fu and S.~J.~Wang,
[arXiv:2310.12932 [astro-ph.CO]].


\bibitem{Giare:2024akf}
W.~Giar\`e,
[arXiv:2404.12779 [astro-ph.CO]].

\bibitem{DiValentino:2018zjj}
E.~Di Valentino, A.~Melchiorri, Y.~Fantaye and A.~Heavens,
Phys. Rev. D \textbf{98} (2018) no.6, 063508
doi:10.1103/PhysRevD.98.063508 [arXiv:1808.09201 [astro-ph.CO]].

\bibitem{Giare:2022rvg}
W.~Giar\`e, F.~Renzi, O.~Mena, E.~Di Valentino and A.~Melchiorri,
Mon. Not. Roy. Astron. Soc. \textbf{521} (2023) no.2, 2911-2918
doi:10.1093/mnras/stad724 [arXiv:2210.09018 [astro-ph.CO]].


\bibitem{Euclid:2024yrr}
Y.~Mellier \textit{et al.} [Euclid],
[arXiv:2405.13491 [astro-ph.CO]].





    \end{thebibliography}
\end{document}